\documentclass{sampta}

\title{Limits of Deterministic Compressed Sensing Considering Arbitrary Orthonormal Basis for Sparsity}

\author{
Arash Amini $^{(1)}$ and Farokh Marvasti $^{(1)}$ \\[12pt]
\addrfnt (1) Advanced Communication Research Institute (ACRI)\\
 EE Department, Sharif University of Technology, Iran.\\  
\email{arashsil@ee.sharif.edu}, \email{marvasti@sharif.edu}
}

\begin{document}
\maketitle

%
% Abstract of the paper
%

\begin{abstract}
It is previously shown that proper random linear samples of a finite discrete signal (vector) which has a sparse representation in an orthonormal basis make it possible (with probability 1) to recover the original signal. Moreover, the choice of the linear samples does not depend on the sparsity domain. In this paper, we will show that the replacement of random linear samples with deterministic functions of the signal (not necessarily linear) will not result in unique reconstruction of $k$-sparse signals except for $k=1$. We will show that there exist deterministic nonlinear sampling functions for unique reconstruction of $1$-sparse signals while deterministic linear samples fail to do so.
\end{abstract}

%
% Series of sections and subsections
%

\section{Introduction}

Recent results in compressed sensing \cite{Baraniuk2007,Candes2006conf,Candes2008,Donoho2006} show that for a class of finite discrete signals named \textit{sparse}, it is possible to reconstruct the signal from significantly lower number of samples than predicted by Nyquist rate in uniform sampling theory. More precisely, if $\mathbf{x}_{n\times 1}$ is a vector whose representation in the orthonormal basis $\mathbf{\Psi}=[\psi_{i,j}]_{n\times n}$ has at most $k$ nonzero elements , it is called $k$-sparse. Also $\mathbf{s}=\mathbf{\Psi}^{H}\cdot \mathbf{x}$ is the translation of $\mathbf{x}$ in the sparsity domain (since $\mathbf{\Psi}$ is orthonormal $\mathbf{\Psi}^{-1}=\mathbf{\Psi}^{H}$); i.e., $\mathbf{s}$ has at most $k$ nonzero elements.

It is proved that if $m$ is of order not less than $O(k\cdot \log(\frac{n}{k}))$, $m$ random linear combination of the samples (multiplication by a random matrix $\mathbf{\Phi}$) of a $k$-sparse signal with length $n$ ($k\ll n$) where the coefficients of the linear samples have normal i.i.d. distribution, suffice (with overwhelming probability) to reconstruct the signal \cite{Donoho2006}. Other distributions have also been proposed which yield the same result \cite{Baraniuk2008}. One of the features of this type of random sampling which was originally of main concern (but no longer as important as it was) is that at the time of sampling it is unimportant in which basis the signal is sparse. The domain of sparsity comes into the picture only at the time of the reconstruction. Therefore, this type of sampling is not matched to a specific sparsity domain. The other advantage of the method lies in its reconstruction method. The mentioned order of $m$ is the same as the order for which the minimization using $\ell_0$ norm for finding the sparse solution could be replaced with $\ell_1$ without considerable change in the solution (solving a polynomial time problem instead of an NP-complete problem) \cite{Candes2005,Gilbert2006conf}. 

Since in most of the applications the input signal has a priory known characteristics, the sparsity domain is also known. Thus, it is logical to design a deterministic sampling matrix ($\mathbf{\Phi}$); in this way, we eliminate the need to save the sampling matrix. Some possible matrices are introduced in \cite{DeVore2007} and \cite{Saligrama2008}.

Although the analog branch of the compressed sensing is being developed parallel to the improvements in the discrete domain \cite{Eldar2008June,Eldar2008Sep}, we will only consider the discrete case in this paper. We will investigate the consequences of deterministic sampling when the sparsity domain is not known at the time of sampling; however, we do not restrict the sampling functions to be linear. In fact we are looking for deterministic functions of $\mathbf{x}$ which uniquely represent all $k$-sparse signals irrespective of their sparsity domain (for reconstruction the domain should be known). 

\section{Preliminaries}
Before we start the main argument, it is necessary to point out a key issue: since the cardinality of $\mathbb{C}$ (or $\mathbb{R}$) is the same as $\mathbb{C}^n$ (or $\mathbb{R}^n$), there exists an injection $g:\mathbb{C}^n\rightarrow \mathbb{C}$ (or $g:\mathbb{R}^n\rightarrow \mathbb{R}$) which as a sampling function, compresses the whole information of a vector into a single sample, even without the assumption of sparsity. Although the compression is amazing, the output sample can not be easily quantized; i.e., the theoretical solution is impractical. For this reason, we are looking for sampling functions which employ the sparsity constraint in order to reconstruct the original signal and then we will discuss how they can be quantized. To formulate the mentioned constraint: we are looking for sampling functions $f_i:\mathbb{C}^n\rightarrow\mathbb{C},~~1\leq i\leq m$ that result in equivalent sampling classes defined by:
\begin{eqnarray}
\mathcal{S}_{s_1,\dots,s_m}=\{\mathbf{x}\in\mathbb{C}^n~~|~~\forall~1\leq i\leq m:~f_i(\mathbf{x})=s_i\}
\end{eqnarray}
such that in each arbitrary orthonormal basis, at most one of the elements of $\mathcal{S}$ has $k$-sparse representation. Moreover, the sparsity constraint should play a key role; i.e., for many cases we should have $|\mathcal{S}|>1$.

First we show why deterministic linear sampling is not a proper choice. Assume the samples are produced as:
\begin{eqnarray}
\left[\begin{array}{c}
s_1\\
\vdots\\
s_m
\end{array}\right]=\left[\begin{array}{c}
f_1(\mathbf{x})\\
\vdots\\
f_m(\mathbf{x})
\end{array}\right]=\mathbf{\Phi}_{m\times n}\cdot \mathbf{x}_{n\times 1}
\end{eqnarray} 

If $m\geq n$, the number of samples exceeds the number of original elements in $\mathbf{x}$, which defeats the underlying purpose of compression. For the case of $m< n$, $\textrm{rank}(\mathbf{\Phi})\leq m<n$ and therefore, the null space of $\mathbf{\Phi}$ will have non-zero dimension:
\begin{eqnarray}
\exists~\mathbf{v}\in\mathbb{C}^n,\|\mathbf{v}\|_{\ell_2}=1:~~\mathbf{\Phi}\cdot\mathbf{v}=0
\end{eqnarray}

Let $\mathcal{V}^{\perp}$ be the subspace of $\mathbb{C}^n$ formed by the vectors perpendicular to $\mathbf{v}$:
\begin{eqnarray}
\mathcal{V}^{\perp}\triangleq\{\mathbf{x}\in\mathbb{C}^n~~|~~\mathbf{v}^H\cdot\mathbf{x}=0\},
\end{eqnarray}
and let $\{\mathbf{u}_1,\dots,\mathbf{u}_{n-1}\}$ be an orthonormal basis for this subspace. Hence, $\{\mathbf{v},\mathbf{u}_1,\dots,\mathbf{u}_{n-1}\}$ forms an orthonormal basis for $\mathbb{C}^n$, or equivalently,
\begin{eqnarray}
\mathbf{\Psi}\triangleq\left[\mathbf{v},~\mathbf{u}_1,~\dots,~\mathbf{u}_{n-1}\right]
\end{eqnarray}
is a unitary matrix. Obviously, the input vector $\mathbf{v}$ is $1$-sparse with respect to the representation in $\mathbf{\Psi}$; however, $\mathbf{v}$ lies in the null space of $\mathbf{\Phi}$ and can not be recovered using its samples (all zero) produced by $\mathbf{\Phi}$.

Thus, for any fixed linear sampling set with less number of samples than the original signal's length, there exists an orthonormal basis and a $1$-sparse signal (in this basis) which can not be uniquely recovered from its samples.

\section{General Sampling for $k$-sparse Signals}
Although deterministic linear sampling fails either in compression or reconstruction, there exists the possibility to introduce general nonlinear sampling functions which do not have these drawbacks. In this section we will show that for $k$-sparse signals with $k\geq 2$ no such sampling functions exist. 

Let us assume that $\{f_i(\mathbf{x})\}_{i=1}^{m}$ are the sampling functions and let $\{s_i\}_{i=1}^{m}$ be the respective samples obtained from a given $k$-sparse vector $\mathbf{x}_0$ (in an orthonormal basis). Reversibility  of the sampling process is equivalent to the fact that $\mathbf{x}_0$ is the only $k$-sparse vector which results in samples $\{s_i\}_{i=1}^{m}$. As previously defined, we represent the sampling equivalent class of the obtained samples $\{s_i\}_{i=1}^{m}$ by $\mathcal{S}$:
\begin{eqnarray}
\mathcal{S}=\big\{\mathbf{x}\in\mathbb{C}^n~\big|~\forall 1\leq i\leq m~,~f_i(\mathbf{x})=s_i \big\}
\end{eqnarray}
Since $\mathbf{x}_0\in S$ we know $\mathcal{S}\neq \emptyset$. Since the case $m\geq n$ contradicts the compression concept, we are assuming $m< n$. Moreover, if for all input vectors $\mathbf{x}_0\in\mathbb{C}^n$ we have $|\mathcal{S}|=1$, sparsity constraint is of no use; i.e., a 1-1 mapping from $\mathbb{C}^n$ to $\mathbb{C}^m$ is formed which is undesired due to its quantization problem. Therefore, there exist input vectors for which the equivalent class $\mathcal{S}$ has at least two elements. We choose $\mathbf{x}_0$ among these input vectors; i.e., there exists $\mathbf{x}_1\neq\mathbf{x}_0$ such that the samples obtained from both of $\mathbf{x}_0$ and $\mathbf{x}_1$ are the same ($\mathbf{x}_0,\mathbf{x}_1\in\mathcal{S}$). Let $\tilde{\mathbf{v}}_0$ be the normal vector in the direction of $\mathbf{x}_0$ and let $\mathbf{v}_1$ be defined as follows:
\begin{eqnarray}
\mathbf{v}_1=\mathbf{x}_1-\langle\mathbf{x}_1,\tilde{\mathbf{v}}_0\rangle\tilde{\mathbf{v}}_0
\end{eqnarray}
where $\langle\mathbf{a},\mathbf{b}\rangle$ stands for the inner product of the vectors $\mathbf{a}$ and $\mathbf{b}$, ($\langle\mathbf{a},\mathbf{b}\rangle=\mathbf{a}^H\cdot\mathbf{b}$). We have two cases:
\begin{enumerate}
\item If $\mathbf{v}_1=\mathbf{0}$, both $\mathbf{x}_0$ and $\mathbf{x}_1$ lie in the direction identified by the normal vector $\tilde{\mathbf{v}}_0$. Similar to the argument in the previous section, there exist normal vectors $\mathbf{u}_1,\dots,\mathbf{u}_{n-1}$ such that $\{\tilde{\mathbf{v}}_0,\mathbf{u}_1,\dots,\mathbf{u}_{n-1}\}$ forms an orthonormal basis for $\mathbb{C}^n$. Obviously both $\mathbf{x}_0$ and $\mathbf{x}_1$ are 1-sparse in this basis and moreover, they produce the same set of samples using the assumed sampling functions. Hence, using the mentioned samples, these two vectors can not be distinguished even with the sparsity constraint.

\item If $\mathbf{v}_1\neq\mathbf{0}$, we define $\tilde{\mathbf{v}}_1$ to be the normal vector in the direction of $\mathbf{v}_1$. Since $\tilde{\mathbf{v}}_0$ and $\mathbf{v}_1$ are orthogonal, $\tilde{\mathbf{v}}_0$ and $\tilde{\mathbf{v}}_1$ are also orthogonal:
\begin{eqnarray}
&&\langle\mathbf{v}_1,\tilde{\mathbf{v}}_0\rangle= \langle\mathbf{x}_1,\tilde{\mathbf{v}}_0\rangle - \langle\mathbf{x}_1,\tilde{\mathbf{v}}_0\rangle \cdot \underbrace{\langle\tilde{\mathbf{v}}_0,\tilde{\mathbf{v}}_0\rangle}_{= \|\tilde{\mathbf{v}}_0\|^2_{\ell_2}=1}=0\nonumber\\
&&~~~~~~~~\Rightarrow~~\mathbf{v}_1 \perp \tilde{\mathbf{v}}_0~~\Rightarrow~~\tilde{\mathbf{v}}_1 \perp \tilde{\mathbf{v}}_0
\end{eqnarray}

Moreover from the definition of $\mathbf{v}_1$ we have:
\begin{eqnarray}
\mathbf{x}_1&=&\mathbf{v}_1+\langle\mathbf{x}_1,\tilde{\mathbf{v}}_0\rangle \tilde{\mathbf{v}}_0 \nonumber\\
&=& \langle\mathbf{x}_1,\tilde{\mathbf{v}}_1\rangle \tilde{\mathbf{v}}_1+\langle\mathbf{x}_1,\tilde{\mathbf{v}}_0\rangle \tilde{\mathbf{v}}_0
\end{eqnarray}
Therefore, $\mathbf{x}_0$ and $\mathbf{x}_1$ are respectively 1-sparse and 2-sparse in any orthonormal basis which contains $\tilde{\mathbf{v}}_0$ and $\tilde{\mathbf{v}}_1$ (existence of such a basis is trivial by considering the subspace of $\mathbb{C}^n$ orthogonal to $\textrm{span}\{\tilde{\mathbf{v}}_0,\tilde{\mathbf{v}}_1\}$). Thus, if $k\geq 2$, both $\mathbf{x}_0$ and $\mathbf{x}_1$ are valid solutions to the sparsity constraint. Thus, again we can not uniquely reconstruct the original signal from its samples.
\end{enumerate}

\section{Sampling for $1$-sparse Signals}
The only case that we are left is the general (nonlinear) sampling of $1$-sparse signals. In fact we will show the existence of such sampling functions for this case, but first lets focus on the conditions that the sampling functions should satisfy:
\newtheorem{lemma}{Lemma}
\begin{lemma}
The set $\mathcal{S}\subset\mathbb{C}^n$ is uniquely 1-sparse decodable iff:
\begin{eqnarray}
\forall ~\mathbf{a}\neq\mathbf{b}\in \mathcal{S}:~~~0<\big|\langle\mathbf{a},\mathbf{b}\rangle\big| < \|\mathbf{a}\|_{\ell_2}\cdot\|\mathbf{b}\|_{\ell_2}
\end{eqnarray}
where $\langle\mathbf{a},\mathbf{b}\rangle$ is the inner product of the two vectors ($\mathbf{a}^H\cdot\mathbf{b}$), $\|.\|_{\ell_2}$ is the $\ell_2$ norm of the vector (square root of the inner product of the vector by itself) and $|.|$ refers to the absolute value of a complex number.
\end{lemma}

Using the Cauchy inequality and the non-negative property of the absolute-value operator, it is straightforward that:
\begin{eqnarray}
\forall ~\mathbf{a},\mathbf{b}\in \mathbb{C}^n:~~~0\leq\big|\langle\mathbf{a},\mathbf{b}\rangle\big| \leq \|\mathbf{a}\|_{\ell_2}\cdot\|\mathbf{b}\|_{\ell_2}
\end{eqnarray}
The only characteristic which distinguishes an arbitrary set from a uniquely 1-sparse decodable set is that the equalities do not occur for the latter. Now we get back to the proof of the lemma. According to the Cauchy theorem, for the right inequality we have:
\begin{eqnarray}
\big|\langle\mathbf{a},\mathbf{b}\rangle\big| = \|\mathbf{a}\|_{\ell_2}\cdot\|\mathbf{b}\|_{\ell_2} ~~\Leftrightarrow~~\exists~c\in\mathbb{C}:~\mathbf{a}=c\mathbf{b}
\end{eqnarray}
If $\mathcal{S}$ contains two unequal vectors $\mathbf{a},\mathbf{b}$ for which the above equality holds, in the orthonormal basis containing the normal vector in the direction of $\mathbf{a}$ (or $\mathbf{b}$) both of the vectors are 1-sparse (since they are in the same direction) which means that $\mathcal{S}$ is not uniquely 1-sparse decodable. On the other hand, if for all unequal vectors of $\mathcal{S}$ strict inequality holds, each line in $\mathbb{C}^n$ passing through the origin can not intersect $\mathcal{S}$ at more than one point. This means that for each vector of an orthonormal basis, there exists at most one element of $\mathcal{S}$ which is 1-sparse in this direction.

The left inequality considers the orthogonality of the vectors in $\mathcal{S}$:
\begin{eqnarray}
\big|\langle\mathbf{a},\mathbf{b}\rangle\big|=0~\Rightarrow~\mathbf{a}\perp\mathbf{b}
\end{eqnarray}
Assume $\mathcal{S}$ contains two nonzero perpendicular vectors $\mathbf{a},\mathbf{b}$ and let $\tilde{\mathbf{a}},\tilde{\mathbf{b}}$ be the normal vectors in their directions, respectively. Since $\tilde{\mathbf{a}}$ and $\tilde{\mathbf{b}}$ are orthogonal, there exists an orthonormal basis for $\mathbb{C}^n$ which contains both $\tilde{\mathbf{a}},\tilde{\mathbf{b}}$; obviously $\mathbf{a}$ and $\mathbf{b}$ are 1-sparse in this basis which implies that $\mathcal{S}$ is not uniquely 1-sparse decodable. 

For the sufficiency of the condition, if no two vectors of $\mathcal{S}$ are perpendicular, among the vectors of each orthonormal basis of $\mathbb{C}^n$ there exists at most one which intersects $\mathcal{S}$ (two intersections reveal a perpendicular pair in $\mathcal{S}$). Also the result of the right-side inequality restricts the number of intersections for each direction to one; so, at most one of the vectors in $\mathcal{S}$ can be 1-sparse in this basis (1-sparse decodablity of $\mathcal{S}$) and the proof of the lemma is complete.

Our next step is to introduce sampling functions whose equivalent sampling classes fulfill the condition of the lemma. We claim that the following three functions always produce such sets:
\begin{eqnarray}
\left\{\begin{array}{l}
f_1(\mathbf{x})=\sum_{i=1}^{n}3^{2(i-1)}\textrm{sign}(\Re\{x_{i}\})\\
~~~~~~~~~~~~~~~~~~~~+3^{2i-1}\textrm{sign}(\Im\{x_{i}\})\\
 \\
f_2(\mathbf{x})=\|\mathbf{x}\|_{\ell_1}=\sum_{i=1}^{n}|x_{i}|\\
 \\
f_3(\mathbf{x})=\sum_{x_i\neq 0}\frac{\textrm{msign}(\Re\{x_i\})~\cdot~\Im\{x_i\} }{|\Re\{x_i\}|+|\Im\{x_i\}|}
\end{array}\right.
\end{eqnarray}
where $\Re\{.\}$ and $\Im\{.\}$ represent the real and imaginary parts, respectively, and $msign(.)$ is the modified sign function which generates the same output as the $sign(.)$ function except that $msign(0)=1$.

The first sampling function, $f_1$, although produces only one sample, uniquely determines the sign of both real and imaginary parts of all the elements of $\mathbf{x}$. Since the output range of $\textrm{sign}$ function is restricted to three different values of $-1$, $0$ and $1$, the sample generated by $f_1$ can be viewed as the base 3 representation of an integer number with the difference that the digit 2 is replaced with -1 (the same residue in division by 3 which yields the uniqueness). Moreover, this sample is an integer between $-\frac{3^{2n}-1}{2}$ and $\frac{3^{2n}-1}{2}$ which requires $\lceil 2n\log_{2}3\rceil$ bits for errorless quantization. We will show that this sample guarantees the left inequality. Let $\mathbf{a}$ and $\mathbf{b}$ be two unequal vectors that $f_1(\mathbf{a})=f_1(\mathbf{b})$, we show that $\langle\mathbf{a},\mathbf{b}\rangle\neq 0$:
\begin{eqnarray}
&&\Re\{\langle\mathbf{a},\mathbf{b}\rangle\}=\Re\{\mathbf{a}^H\cdot\mathbf{b}\}\nonumber\\
&&~~~~~~~~~~~=\sum_{i=1}^{n}\Re\{a_i\}\Re\{b_i\}+\Im\{a_i\}\Im\{b_i\}
\end{eqnarray}
Since $f_1(\mathbf{a})=f_1(\mathbf{b})$, we have $sign(\Re\{a_i\})=sign(\Re\{b_i\})$ and $sign(\Im\{a_i\})=sign(\Im\{b_i\})$ for all $1\leq i \leq n$. Therefore $\Re\{a_i\}\cdot\Re\{b_i\}+\Im\{a_i\}\cdot\Im\{b_i\}\geq 0$ and the equality occurs if and only if $\Re\{a_i\}=\Re\{b_i\}=\Im\{a_i\}=\Im\{b_i\}=0$. Consequently we have:
\begin{eqnarray}
&&\sum_{i=1}^{n}\Re\{a_i\}\cdot\Re\{b_i\}+\Im\{a_i\}\cdot\Im\{b_i\}\geq 0 \nonumber\\
&&\Rightarrow \Re\{\langle\mathbf{a},\mathbf{b}\rangle\}\geq 0
\end{eqnarray}
with the equality for $\mathbf{a}=\mathbf{b}=\mathbf{0}$ which contradicts the fact that $\mathbf{a}$ and $\mathbf{b}$ are unequal. Thus:
\begin{eqnarray}
\left\{\begin{array}{l}
\mathbf{a}\neq\mathbf{b} \\
f_1(\mathbf{a})=f_1(\mathbf{b})\\
\end{array}\right.
&\Rightarrow&\Re\{\langle\mathbf{a},\mathbf{b}\rangle\}> 0 \nonumber\\
&\Rightarrow& \big|\langle\mathbf{a},\mathbf{b}\rangle\big| > 0
\end{eqnarray}

Now using the samples generated by $f_2$ and $f_3$ (in addition to $f_1$) we prove the right inequality. If $\mathbf{a}$ and $\mathbf{b}$ are such that $f_2(\mathbf{a})=f_2(\mathbf{b})$, $f_3(\mathbf{a})=f_3(\mathbf{b})$ and $\big|\langle\mathbf{a},\mathbf{b}\rangle\big|=\|\mathbf{a}\|_{\ell_2}\cdot \|\mathbf{b}\|_{\ell_2}$, $\mathbf{a}$ and $\mathbf{b}$ must have the same direction (Cauchy inequality) or equivalently, there exists $c\in\mathbb{C}$ such that $\mathbf{b}=c~\mathbf{a}$:
\begin{eqnarray}
f_2(\mathbf{b})=f_2(c~\mathbf{a})=\sum_{i=1}^n\big|c\cdot a_i\big|&=&|c|\sum_{i=1}^n |a_i|\nonumber\\
&=&|c|f_2(\mathbf{a})
\end{eqnarray}
The condition $f_2(\mathbf{a})=f_2(\mathbf{b})$ restricts the amplitude of $c$ to have the value 1 ($|c|=1$); note that $f_2(\mathbf{a})=0$ results in $\mathbf{a}=\mathbf{b}=0$ which again contradicts the previous assumption of distinctness of $\mathbf{a}$ and $\mathbf{b}$. Thus, up to this point we shown that by having the first two samples (generated by $f_1$ and $f_2$), the original vector within a constant complex coefficient on the unit circle in known. We show that the third sample uniquely determines the phase of that coefficient which implies the overall uniqueness. Before we proceed to reveal the role of the third sample, let us consider the effect of quantizing the second sample. Since the absolute-value operator is a continuous function, the sampling function $f_2$ is also a continuous function of the input vector. Continuity implies that small perturbations in the input such as quantization are mapped to small deviations in the output. Moreover, the quantization of this sample does not affect the mentioned restriction on $c$ ($|c|=1$); i.e., the ambiguity is still the phase of $c$. In simple words, although the reconstructed amplitudes do not exactly match the original ones, the uniqueness (in the amplitude) is still valid. In fact, due to quantization, the reproduced amplitudes fall within a precision from the original values. If the input elements of the vector are quantized with $l$ bits prior to sampling, the output of the $f_2$ can be quantized with  $l+\lceil \log_2(n) \rceil$ bits (unsigned) without causing any further loss of information.

Now we show the role of the last sample in revealing the phase of the nonzero elements of the input vector in the sparsity domain. Since the summation in $f_3$ is taken over the nonzero elements of the vector, the denominator in each of the terms is a positive real number (the denominator becomes zero only when $\Re\{x_i\}=\Im\{x_i\}=0$ which is excluded by $x_i\neq 0$). Besides, we have:
\begin{eqnarray}
0\leq&&\bigg|\frac{\textrm{msign}(\Re\{x_i\})~\cdot~\Im\{x_i\}}{|\Re\{x_i\}|+|\Im\{x_i\}|}\bigg|\nonumber\\
&=&\frac{|\Im\{x_i\}|}{|\Re\{x_i\}|+|\Im\{x_i\}|}~~~\leq 1
\end{eqnarray}
which shows that sample generated by $f_3$ is bounded between $-n$ and $n$ (boundedness is necessary for quantization).

If $\mathbf{a}$ and $\mathbf{b}$ are two vectors which produce the same set of samples (all three), we know $\mathbf{b}=c~\mathbf{a}$ where $|c|=1$ by use of the first two samples. Considering the third sample we should have $f_3(\mathbf{a})=f_3(\mathbf{b})$:
\begin{eqnarray}
&&\sum_{a_i\neq 0} \frac{\textrm{msign}(\Re\{a_i\})~\cdot~\Im\{a_i\}}{|\Re\{a_i\}|+|\Im\{a_i\}|} \nonumber\\
&=& \sum_{c a_i\neq 0} \frac{\textrm{msign}(\Re\{ca_i\})~\cdot~\Im\{ca_i\}}{|\Re\{ca_i\}|+|\Im\{ca_i\}|}
\end{eqnarray}
Owing to the fact that $|c|=1$, the conditions $a_i\neq 0$ and $ca_i\neq 0$ are equivalent; therefore, the terms used in the summations are the same in both sides. In addition, the first sample uniquely determines the $sign$ and therefore  $msign$ of both real and imaginary parts; thus, we should have $msign(\Re\{a_i\})=msign(\Re\{ca_i\})$ and $sign(\Im\{a_i\})=sign(\Im\{ca_i\})$. To simplify the notations, we represent the real and imaginary parts of $a_i$ and $ca_i$ by $R_i,I_i$ and $R_i^{'},I_i^{'}$, respectively. Moreover, by $sn$ and $\tilde{sn}$ we mean the $sign$ and $msign$ operators:
\begin{eqnarray}
&&\frac{\tilde{sn}(R_i)~\cdot~I_i}{|R_i|+|I_i|} - \frac{\tilde{sn}(R_i^{'})~\cdot~I_i^{'}}{|R_i^{'}|+|I_i^{'}|} \nonumber\\
&&=\tilde{sn}(R_i)\cdot sn(I_i)\bigg(\frac{|I_i|}{|R_i|+|I_i|} - \frac{|I_i^{'}|}{|R_i^{'}|+|I_i^{'}|}\bigg)\nonumber\\
&&=\tilde{sn}(R_i)\cdot sn(I_i)\frac{|I_i|.|R_i^{'}|- |I_i^{'}|.|R_i|}{\big(|R_i|+|I_i|\big)\big(|R_i^{'}| +|I_i^{'}|\big)}\nonumber\\
&&=\frac{I_i.R_i^{'}- I_i^{'}.R_i}{\big(|R_i|+|I_i|\big)\big(|R_i^{'}| +|I_i^{'}|\big)}\nonumber\\
&&=\frac{-\Im\{c\}\cdot|a_i|^2}{\big(|R_i|+|I_i|\big)\big(|R_i^{'}| +|I_i^{'}|\big)}
\end{eqnarray}
Thus, $f_3(\mathbf{a})=f_3(\mathbf{b})$ yields:
\begin{eqnarray}
&&-\Im\{c\}\underbrace{\sum_{a_i\neq 0}\frac{|a_i|^2}{\big(|R_i|+|I_i|\big)\big(|R_i^{'}| +|I_i^{'}|\big)}}_{>~0}=0\nonumber\\
&&~~~~~~~~~~~~~~~~~~~\Rightarrow~~~\Im\{c\}=0
\end{eqnarray}
The above result in conjunction with the previous condition that $|c|=1$ remains two possible choices for $c$, $1$ or $-1$; nonetheless, $c=-1$ produces different signs for real and imaginary parts in $\mathbf{b}$ which is not acceptable by the choice of the first sample. As a consequence, $c=1$ is the only choice which means that $\mathbf{a}=\mathbf{b}$ and the proof for the claimed unique 1-sparse decodability of the aforementioned set of sampling functions is complete. A similar argument to the one presented for quantization of the second sample is valid here; even after quantization, there exists at most one 1-sparse vector which produces these samples. Given that the quantization does not change the sign of the real and imaginary parts, the sampling functions will still be in their continuous region and therefore, small errors in the input (such as quantization) will be mapped to small errors in the reconstructed signal. 

It is easy to check that if the domain is $\mathbb{R}$ rather than $\mathbb{C}$, the third sample is not required; i.e., the sign of the elements and the $\ell_1$ norm of the vector suffice for unique representation of 1-sparse vectors.

\section{Conclusions}

We have considered the deterministic sampling of sparse signals whose sparsity domain is not known at the time of sampling. Although random linear sampling of such signals has been shown to be a proper choice (under some conditions), the deterministic approach fails in reconstruction of some identifiable subclasses. The mentioned drawback is still an issue when nonlinear measurements of the $k$-sparse signals where $k>1$ are employed. It is shown that class of $1$-sparse signals (in an arbitrary linear sparsity domain) can be uniquely identified using nonlinear sampling. We have demonstrated a necessary and sufficient condition for the sampling functions to provide the unique 1-sparse decodability of the samples, in addition to presenting a realizable set of such functions.

%
% Bibliography
%

\bibliographystyle{plain}
\bibliography{sampta}  

% File sampta.bib should exist 
% Process latex bibtex latex latex to resolve all references
%

% end of sampta.tex file
\end{document}